\documentclass{ws-mpla}
\usepackage[super]{cite}
\usepackage[utf8]{inputenc}
\usepackage[]{cite}
\usepackage{amsmath}
\usepackage{amsfonts}
\usepackage[breaklinks=true]{hyperref}
\usepackage{color}
\usepackage[final]{graphicx}
\usepackage{subcaption}
\begin{document}

\markboth{Authors' names}{Interacting and Non-interacting R\'{e}nyi Holographic Dark Energy Models in DGP Braneworld
}

%%%%%%%%%%%%%%%%%%%%% Publisher's Area please ignore %%%%%%%%%%%%%%
\catchline{}{}{}{}{}
%%%%%%%%%%%%%%%%%%%%%%%%%%%%%%%%%%%%%%%%%%%%%%%%%%%%%%%%%%%%%%%%%%%

\title{Interacting and Non-interacting R\'{e}nyi Holographic Dark Energy Models in DGP Braneworld
}

\author{Arindam Saha}

\address{Jalpaiguri Govornment Engineering College\\
Jalpaiguri, West Bengal, India 735102\\
arindamjal@gmail.com}

\author{Anirban Chanda, Sagar De}

\address{Department of Physics\\ University of North Bengal\\ Rajarammohunpur, Darjeeling, West Bengal, India 734013\\
anirbanchanda93@gmail.com
}

\author{Souvik Ghose}
\address{Harishchandra Research Institute\\
Prayagraj(Allahabad), UP, India 211019\\
souvikghose@hri.res.in}
\author{B. C. Paul}

\address{Department of Physics\\ University of North Bengal\\ Rajarammohunpur, Darjeeling, West Bengal, India 734013\\
anirbanchanda93@gmail.com
}

\maketitle

\pub{Received (Day Month Year)}{Revised (Day Month Year)}

\begin{abstract}
We investigate both the interacting and non-interacting R\'{e}nyi Holographic Dark Energy (RHDE) models in DGP brane world framework. Cosmological parameters and their evolutions are probed to obtain realistic cosmological models. We note that both the models accommodate the present accelerating phase of expansion with the observed dark energy density. Classical stability of the cosmological model and $Om$- diagnostic are also studied to test the suitability of the cosmological models obtained in the presence of RHDE in DGP braneworld.

\end{abstract}

\ccode{PACS Nos.: include PACS Nos.}

\section{Introduction}
Cosmological models based on the general theory of relativity (GR)  can not naturally include the current accelerating expansion of the universe \cite{riess,perlm,perlm2,perlm3}. Many interesting modifications of the underlying theory of gravitation \cite{carroll, uddin, linderfg, samipaddy, gibb, nojde1, cappode, liucamp} thus surfaced over the years. However, till now, the simplest way to induce acceleration is to consider a constant vacuum energy term in the Einstein Hilbert action which is the famous cosmological constant. The cosmological constant, if that is propelling the present phase of acceleration of the universe, has an astonishingly small value compared to anything that can be calculated from the theory \cite{weinberg1989cosmological}. This discord initially inspired the use of the Holographic principle in cosmology which would place an upper bound in the calculation of vacuum energy density from the field theory. The principle states that the entropy of the system grows with the area rather than the volume \cite{susskind,cohen,lii}. Thus, the conventional field theories were overestimating vacuum energy density due to their consideration of additional degrees of freedom. The vacuum energy evolves with time obeying the Holographic principle.
\begin{equation}
\label{eq:hol-1}
\rho_{\Lambda}=\frac{3c^2M_P^2}{L^2},
\end{equation}
where, $\rho_\Lambda$ is related to the ultraviolet (UV) cutoff, $L$ denotes the infrared cutoff, $M_P$ is the Planck mass and $c$ is a numerical constant. When applied to cosmology the holographic principle does not suggest any particular choice of an infra-red cutoff. The last few years have seen an increased interest in this field and different options, such as the particle horizon, the future event horizon, and the Hubble horizon, have been studied in detail (see \cite{miao2011} for a review). Also, non-extensive probability distribution has been considered in many recent literatures. Consequently, R\'{e}nyi, Tsallis \cite{tsalis,renyi1970probability}  and other generalized forms of entropy \cite{nojiri2006unifying,nojiri2022barrow,nojiri2021different,nojiri2017covariant} have been discussed. In the correct limit, these forms yield Bekenstein’s entropy. For example, RHDE can be written as \cite{morad1}:
\begin{equation}
\label{eq:rhde1}
\rho_\Lambda = \frac{3c^2M_P^2}{L^2}(\pi \delta L^2 +1)^{-1}
\end{equation}
which reverts to normal Holographic dark energy when $\delta$ vanishes.
On the other hand, recent advances in high energy physics, leading to the unification of the other three forces of nature into a single theory, have encouraged higher dimensional theories where people have seen the possibility of unifying gravity with the other gauge theories \cite{kaluza,klein}. Braneworld scenario is an approach to this direction and there are at least three main discourses of this approach: (i) the Randall-Sundrum (RS II) brane model \cite{randall1999large,randall1999alternative}, (ii) Dvali-Gabadadze-Porrati (DGP) braneworld \cite{dvali20004d}, and (iii) and the cyclic model of Steinhardt and Turok \cite{steinhardt2002cosmic, khoury2004designing}. In the present study, we shall concentrate on the second one.\\
DGP braneworld model considers our four-dimensional universe as a hyper-surface (brane) in a five-dimensional Minkowski universe (bulk). While gravity can propagate to the bulk \cite{randall1999large, dvali20004d}, the other gauge forces are confined to the four-dimensional brane only. However, there exists a cross-over scale ($r_c$)below which gravity is also four-dimensional. This scale also induces the infrared modification of gravity.  Above the cross-over scale, gravity can leak into the five-dimensional bulk. In the cosmological context, this leakage generates the present phase of acceleration of the universe \cite{ghaffari2019tsallis}. The DGP model has two branches of development. The self-accelerating branch can produce the late-time acceleration without any dark energy but suffers from the ghost degrees of freedom in the linearized theory and is in conflict with the observation \cite{koyama2007ghosts}. The normal branch, on the other hand, is free from such issues but requires dark energy, at least in the form of a cosmological constant, to generate any late-time acceleration of the universe. We shall consider this branch for our study. In this work, we assume that like normal HDE equation in eq.(\ref{eq:hol-1}) (see \cite{setare2009holographic,wu2008dynamics} RHDE correspondence equation (eq. (\ref{eq:rhde1})) also holds in DGP model and consider the evolution of dark energy according to Holographic principle. The assumption is also justified as the Holographic principle is applicable in any dimension \cite{bousso1999holography,iwashita2006holographic}. \\
The cosmological implication of Holographic Dark Energy (HDE hereafter) in the DGP brane framework has been studied in \cite{iqbal2019tsallis,wu2008dynamics}. Recently Setare has explored the Chaplygin gas correspondence of HDE in DGP braneworld \cite{setare2009holographic}. A modified HDE model has been explored in the same framework by Liu, Wang and Yang \cite{liu2010modified}. However, to date, scant attention has been paid to the detailed construction of Re\'nyi Holographic dark energy models (RHDE) in DGP. The embedding of higher dimensional space in DGP permits the consideration of brane or bulk matter. Here we consider the matter and dark energy components and investigate the HDE correspondence in both interacting and non-interacting scenarios and study the evolution of important cosmological parameters. The finding indicates that the dark energy might have evolved from a past phantom phase.
The present study is divided into six different sections. In the next sect., we briefly discuss the idea of R\'{e}nyi Holographic dark energy. Field equations of the DGP brane are derived in sect. \ref{feq}. Non-interacting and interacting RHDE models in the DGP brane are presented in sect. \ref{rhdeni} and \ref{rhdeint}. In sect. (\ref{clstable}) the classical stability of the models is discussed. Two different diagonistcs for the model are given in sect. \ref{sec:diagn}. Finally, in sec. \ref{disc} findings of the present study are discussed.

%--------------------------------Basics of Re\'nyi------------------------------------------------
\section{Basics of Re\'{n}yi Holographic dark energy}
\label{renbr}
Non-extensive generalization thermodynamics defines Tsallis entropy ($S_T$) \cite{tsalis,renyi1970probability,biro2013q,czinner2016renyi,belin2013holographic} for $W$ as: 
\begin{equation}
\label{tsaen}
S_{T}=k_{B} \frac{1- \sum_{i=1}^{W}p_{i}^{q}}{q-1} \; \; \left(\sum_{i=1}^{W}p_{i}=1; \; q\in \rm I\!R \right),
\end{equation}
where, the $i^{th}$ microstate has an associated probability $p_i$ ($\sum_{i}p_{i}=1$) and $q$ is a real number. Cosmology of the Tsallis holographic dark energy has been the focus of a number of works \cite{sharma2020diagnosing,varshney2019statefinder,thdemain}. Originally, R\' enyi entropy ($S_{R}^{Org}$), is given as \cite{renyi1970probability,komatsu2017cosmological}:
\begin{equation}
\label{rsaen}
S_{R}^{Org}=k_{B} \frac{ln\sum_{i=1}^{W}p_{i}^{q}}{q-1}=\frac{1}{1-q}ln\left[1+(1-q)S_{T}\right].
\end{equation}
Eq.(\ref{tsaen}) and eq.(\ref{rsaen}) both lead to Boltzmann-Gibbs entropy for $q=1$. Bekenstein-Hawking ($S_{BH}$) entropy can also be thought as a kind of non-extensive entropy which leading to a R\' enyi entropy \cite{czinner2016renyi, morad1}:
\begin{equation}
\label{rsaen1}
S_{R}=\frac{1}{\delta}ln\left(1+\delta S_{BH} \right),
\end{equation}
where $\delta=1-q$ and  $S_{R}=S_{BH}$ for $\delta=0$.
 RHDE based cosmology has received considerable attention of late \cite{sharma2020statefinder,dubey2020diagnosing} for a detailed discussion on R\'enyi entropy, see \cite{morad1}) and is discussed in the present work considering Hubble horizon as the infrared cut off. The dark energy correspondence equation is given by eq.(\ref{eq:rhde1}). 
%---------------------------------------------- Field Equations ---------------------------------------
\section{Field equations of DGP Braneworld}
\label{feq}

For spatially flat universe, the modified  Friedmann equations on the brane can be written for DGP model \cite{liu2010modified}:
\begin{equation}
\label{fried-dgp}
H^2 = \left( \sqrt{\frac{\rho}{3M_P^2}+\frac{1}{4r_c^2}}+\frac{\epsilon}{2r_c}\right),
\end{equation}
where $H=\frac{\dot{a}}{a}$ is the Hubble constant, $a$ is the scale factor and $r_c$ is the crossover scale. The DGP model gives the usual 4-dimensional Einstein's theory for $r<<r_c$. The two branches of DGP are represented by $\epsilon = \pm 1$. As mentioned previously,$\epsilon = -1$ gives a self accelerating universe. For this work, we consider $\epsilon = 1$.
Assuming $M_{pl}=1$ the eq. (\ref{fried-dgp}) can be re-written as:
\begin{equation}
\label{fried-loop}
    H^{2}-\epsilon\frac{H}{r_{c}}=\frac{\rho_{m}+\rho_{D}}{3}.
\end{equation}
Here we denote the total energy density by $\rho = \rho_{m} + \rho_{D}$ where $\rho_{m}$ is the density of matter (including dark matter and normal baryonic matter), and $\rho_{D}$ represents the dark energy density. In absence of matter field ($\rho=0$) considering $\epsilon=1$, eq. (\ref{fried-loop}) can be expressed as:
\begin{equation}
\label{fried-rc}
    H=\frac{1}{r_{c}},
\end{equation}
leading to a solution $a(t)=a_{0}e^{\frac{t}{r_{c}}}$. This solution describes an accelerating universe with a constant EoS parameter $w_{D}=-1$ which resembles the cosmological constant. The solution, however, suffers from the same fine tuning problem and the cosmic coincidence problem like any other model with a cosmological constant. The dark energy may also be dynamical in nature> Observations do not rule out the possibility that dark energy might have made a transition from $w_{D}>-1$ to $w_{D}<-1$  at some earlier epoch. Due to these issues and the fact that to arrive at eq. (\ref{fried-rc}) we have ignored all parts of energy on the brane including DE, dark matter and baryonic matter which is not a valid assumption, we have considered $\epsilon=-1$ to study the evolution of the universe.\\
Considering no interaction between the matter and dark energy components, and no energy exchange between the brane and the bulk, the energy conservation equations can be written as:
\begin{equation}
\label{cons-ni1}
    \dot{\rho}_{m}+3H\rho_{m}=0 \Rightarrow \rho_{m}=\rho_{0}(1+z)^{3},
\end{equation}
\begin{equation}
\label{cons-ni2}
    \dot{\rho}_{D}+3H(1+w_{D})\rho_{D}=0,
\end{equation}
where, $\rho_{0}$ is a constant of integration and we have expressed the scale factor ($a$) in terms of the redshift parameter ($z$) as $a=\frac{1}{1+z}$ with the current value of $a$ being normalized to one. We define the dimensionless density parameters as,
\begin{equation}
\label{omegas}
\begin{aligned}
    \Omega_{m}=\frac{\rho_{m}}{\rho_{cr}}=\frac{\rho_{m}}{3M_{Pl}^{2}H^{2}}; \\
    \Omega_{D}=\frac{\rho_{D}}{\rho_{cr}}=\frac{\rho_{D}}{3M_{pl}^{2}H^{2}};\\
		\Omega_{rc}=\frac{1}{4H^{2}r_{c}^{2}}.
		\end{aligned}
\end{equation}
Using the above relations one can express eq.(\ref{fried-loop}) as:
\begin{equation}
\label{omeqone}
    \Omega_{m}+\Omega_{D}+2\epsilon \sqrt{\Omega_{rc}}=1.
\end{equation}

%%%%%%%%%%%%%%%%%%%%%%
For $\epsilon=-1$ the eq. (\ref{omeqone}) can be rewritten as:
\begin{equation}
\label{omegarc}
\Omega_{m}+\Omega_{D}=1+2\sqrt{\Omega_{rc}}.
\end{equation}
%%%%%%%%%%%%%%%%%%%%%
Taking the time derivative of eq.(\ref{fried-loop}) yields:
\begin{equation}
\label{hdoth-ni}
     2\dot{H}H+\frac{\dot{H}}{r_{c}}=\frac{1}{3}(\dot{\rho}_{m}+\dot{\rho}_D).
\end{equation}
%%%%%%%%%%%%%%%%%%%%
Using the conservation equations (eq. \ref{cons-ni1} and eq. (\ref{cons-ni2})) we finally obtain the relation,
\begin{equation}
\label{hdot-ni}
    \frac{\dot{H}}{H^{2}}=-\frac{3}{2}\frac{1+2\dot{\Omega}_{rc}+2_{D}\Omega_{D}}{1+\sqrt{\Omega_{rc}}}.
\end{equation}
%%%%%%%%%%%%%%%%%%%%
%*****************************************%

%******************************************%
\section{RHDE in DGP Braneworld with non-interacting fluids}
\label{rhdeni}

%**********************************************
\begin{figure}[ht]
	\centering	
	\begin{subfigure}{0.45\textwidth}
		\centering
		\includegraphics[width = 0.8 \linewidth]{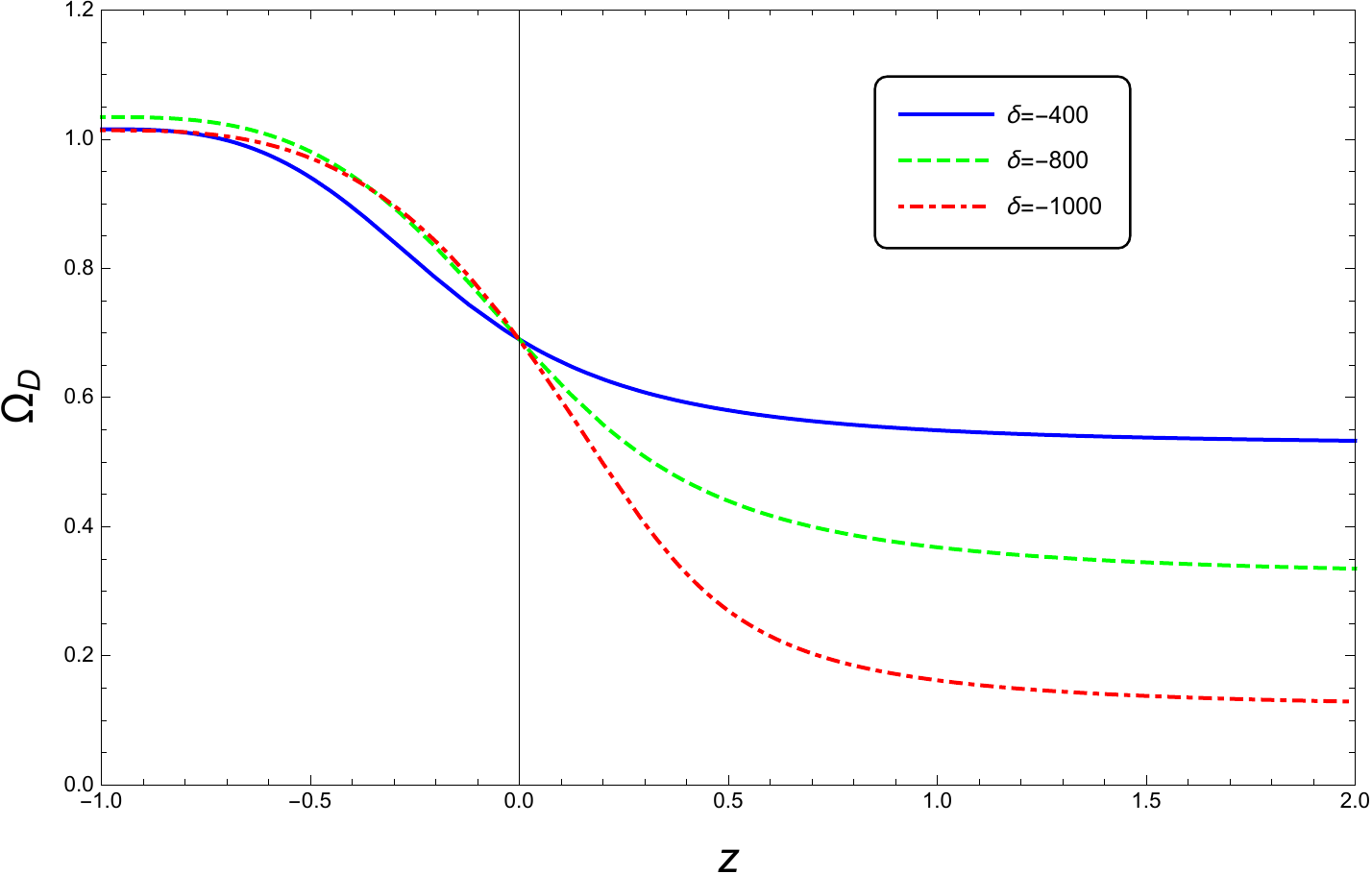}
		\caption{$\Omega_{D}$ vs $z$ for $\Omega_{D0}=0.69$ and different $\delta$ values.}
		\label{fig:omz-ni}
	\end{subfigure}
	\begin{subfigure}{0.45\textwidth}
		\centering
		\includegraphics[width = 0.8 \linewidth]{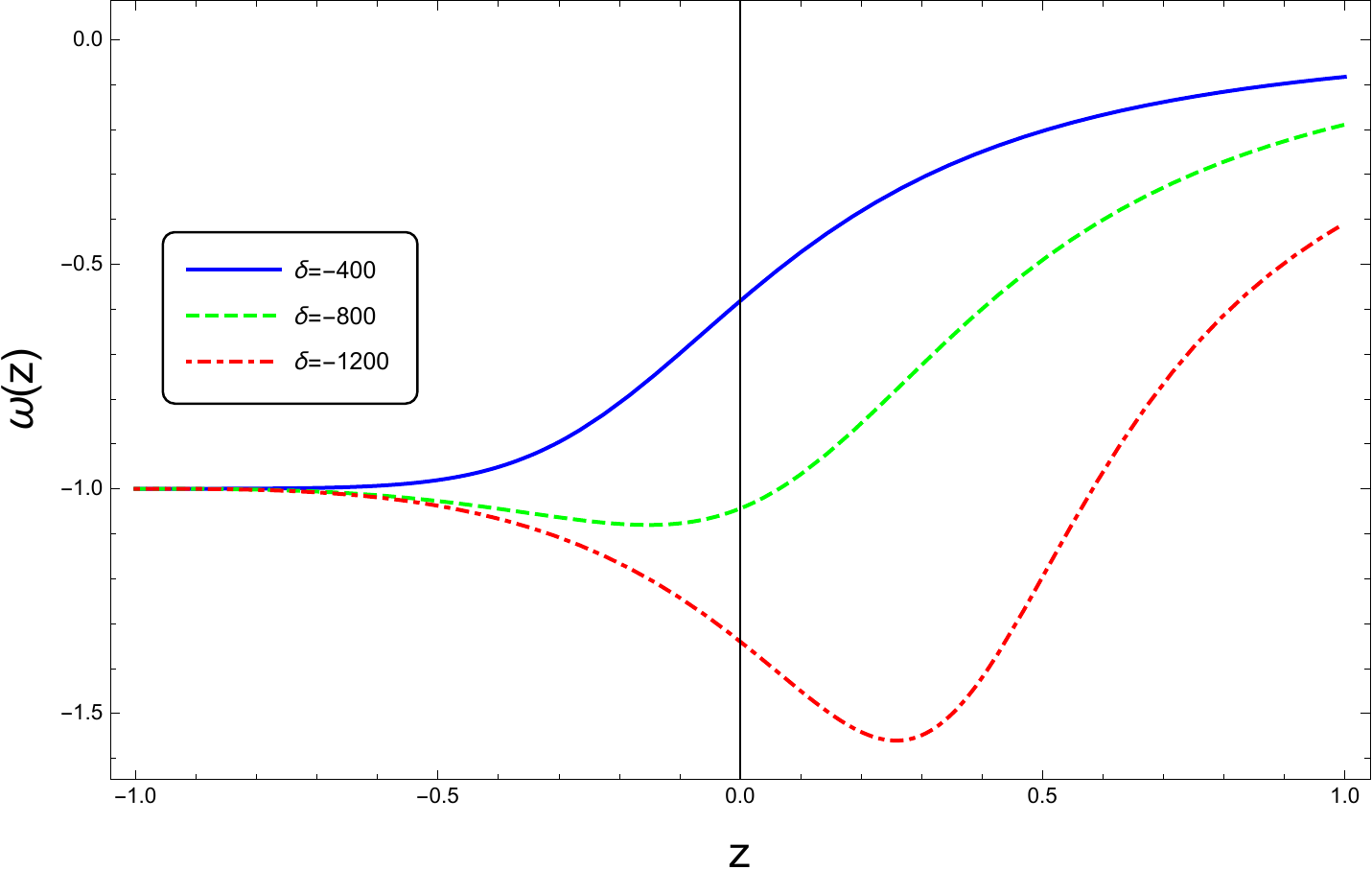}
		\caption{$w_{D}$ vs $z$ for $\Omega_{D0}=0.69$ and different $\delta$ values.}
		\label{fig:eosz-ni}
	\end{subfigure}
	\begin{subfigure}{0.45\textwidth}
		\centering
		\includegraphics[width = 0.8 \linewidth]{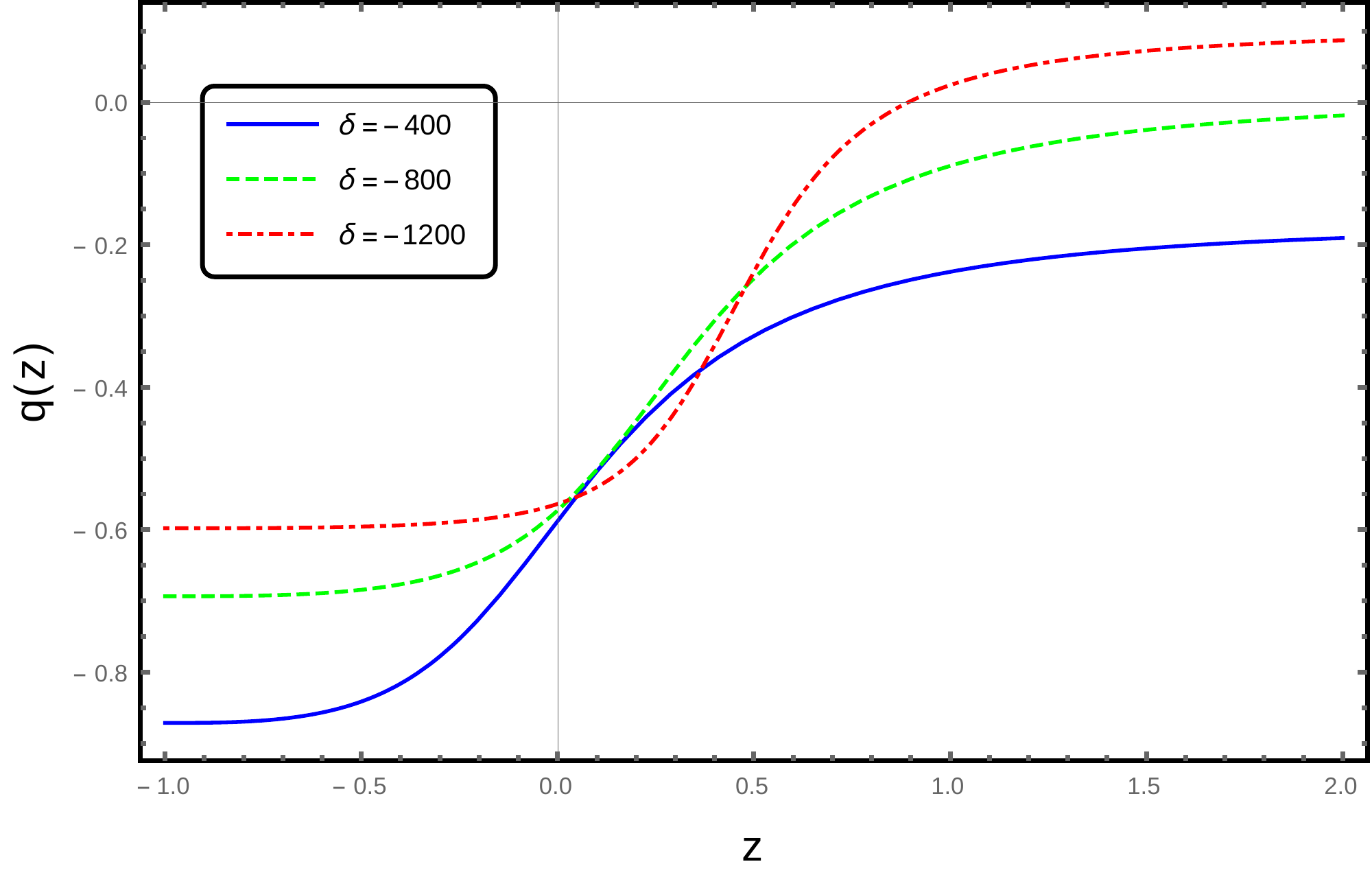}
		\caption{$q(z)$ vs $z$ for $\Omega_{D0}=0.69$ and different $\delta$ values.}
		\label{fig:decz-ni}
	\end{subfigure}
	\begin{subfigure}{0.45\textwidth}
		\centering
		\includegraphics[width = 0.8 \linewidth]{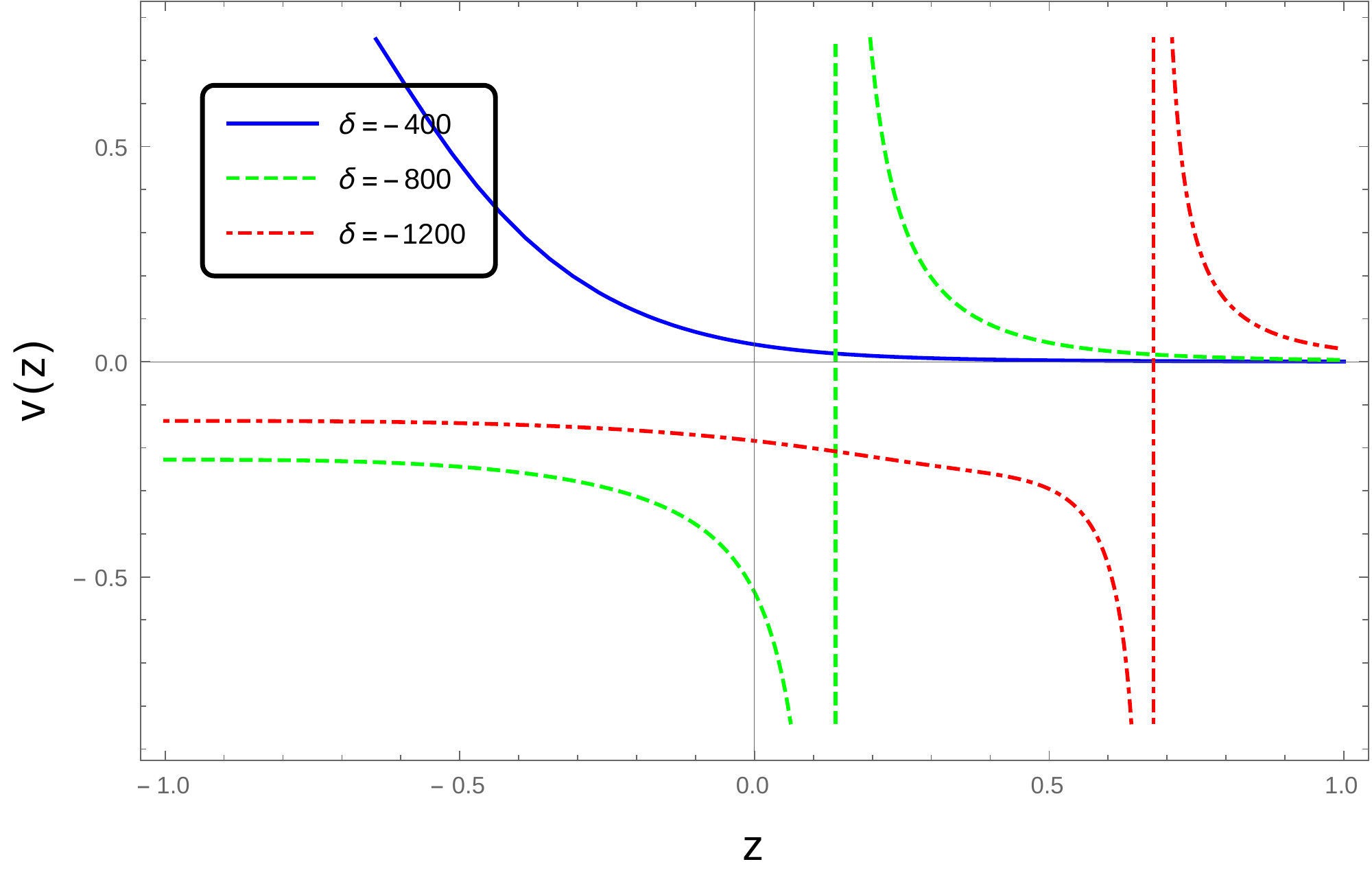}
		\caption{$v^{2}$ vs $z$ for $\Omega_{D0}=0.69$ and different $\delta$ values.}
		\label{fig:vsq-ni}
	\end{subfigure}
	\label{test1}
	\caption{Evolution of cosmological parameters and stability of the non-interacting RHDE model in DGP brane.}    	
\end{figure}
%***********************

Considering RHDE with the Hubble radius as IR cut off ($L=\frac{1}{H}$), the dark energy density can be expressed as \cite{morad1},
\begin{equation}
\label{rhod-ni}
    \rho_{D}=BH^{2}(1+\frac{\delta \pi}{H^{2}})^{-1}.
\end{equation}
%%%%%%%%%%%%%%%%%%%%
where, $B=\frac{3C^{2}}{8\pi}$, and $C^{2}$ is a numerical constant. The corresponding dark energy density parameter can be expressed as $\Omega_{D}=\frac{c^{2}}{1+\frac{\delta \pi}{H^{2}}}$. The Hubble parameter can be expressed in terms of the dark energy density parameter as:
\begin{equation}
\label{hub-ni}
    H^{2}=\frac{\delta\pi\Omega_{D}}{C^{2}-\Omega_{D}}.
\end{equation}
%%%%%%%%%%%%%%%%%%%%
Taking the time derivative of equation eq. (\ref{hub-ni}) and changing the variable we get,
\begin{equation}
\label{omprime-ni}
    \Omega'_{D}=\frac{2(C^{2}-\Omega_{D})\Omega_{D}}{C^{2}}\frac{\dot{H}^{2}}{H^{2}},
\end{equation}
%##################
where prime denotes the derivative respect to $x=lna$, and we have used the relation $\dot{\Omega_{D}}=H\Omega_{D}'$ to write eq. (\ref{omprime-ni}). Fig. (\ref{fig:omz-ni}) illustrates the evolution of $\Omega_D$ with redshift. In plotting the figures, we have used $C = 0.818^{+0.113}_{-0.097}$ as indicated by observation in normal Holographic Dark energy scenario. That way, RHDE should correctly yield HDE as a limiting condition \cite{Li:2009bn}. in the limiting situation The non-interacting model successfully incorporates the present day value of $\Omega_D$ which is independent of model parameters. The dark energy equation of state (EoS hereafter) parameter can be obtained from eq. (\ref{cons-ni2}) as:
\begin{equation}
\label{eq:omd1}
    w_{D}=-1-\frac{\dot{\rho_{D}}}{3H\rho_{D}}.
\end{equation}
For RHDE, using eq. (\ref{rhod-ni}) we rewrite the above equation as:
\begin{equation}
\label{eq:omd2}
    w_{D}=\frac{C^{2}(1+3\sqrt{\Omega_{rc}})-\Omega_{D}(1+2\sqrt{\Omega_{rc}})}{C^{2}(1+\sqrt{\Omega_{rc}})-(2C^{2}-\Omega_{D})\Omega_{D}}.
\end{equation}
%%%%%%%%%%%%%%%%%%%%%%%%

Variation of the EoS parameter with redshift is shown in fig.(\ref{fig:eosz-ni}) for different $\delta$ values. The only theoretical constraint on the $\delta$ values comes from the entropy equation. The entropy of a black-hole should remain positive irrespective of the value of $\delta$ and we checked this consistency \cite{Dubey:2021lmm}. Such values were considered in earlier literature as well \cite{morad1, Dubey:2021lmm}. Note that the EoS parameter stays negative in this case throughout the evolution of the universe. There is also a transition from quintessence to the phantom domain for a positive redshift. For the $\delta$ values considered here, DE behaves as a phantom fluid and transits into the quintessence era in future. For smaller $\delta$ values the EoS parameter dips further into the phantom domain. For larger values of delta, the dark energy never enters any phantom phase. However, as we shall see later that a higher $\delta$ value has its trade-off.\\
The deceleration parameter $(q=-1-\frac{\dot{H}}{H^{2}})$ in this case can be expressed as:
\begin{equation}
\label{dec-ni}
\begin{aligned}
q= & \frac{1}{2(1+\sqrt{\Omega_{rc}})[C^{2}(1+\sqrt{\Omega_{rc}})-(2C^{2}-\Omega_{D})\Omega_{D}]}\\
   & \left[(5C^{2}-C^{2}\Omega_{D}-2\Omega_{D}^{2})\sqrt{\Omega_{rc}}+\right.\\
	 & \left. C^{2}(1+4\Omega_{rc})+\Omega_{D}(C^{2}-\Omega_{D})-3\Omega_{D}^{2}\right],
\end{aligned}
\end{equation}
The variation of the deceleration parameter is illustrated in fig.(\ref{fig:decz-ni}). The universe evolves into the present accelerating phase from a decelerating phase in the past. This point of transition depends on the model parameter $\delta$ and for relatively larger values the transition may never occur at all. Note that an eternally accelerating universe posses a challenging scenario for structure formation.\\

%***************************
\begin{figure}
	\centering
	\begin{subfigure}{0.45\textwidth}
		\centering
		\includegraphics[width = 0.8 \linewidth]{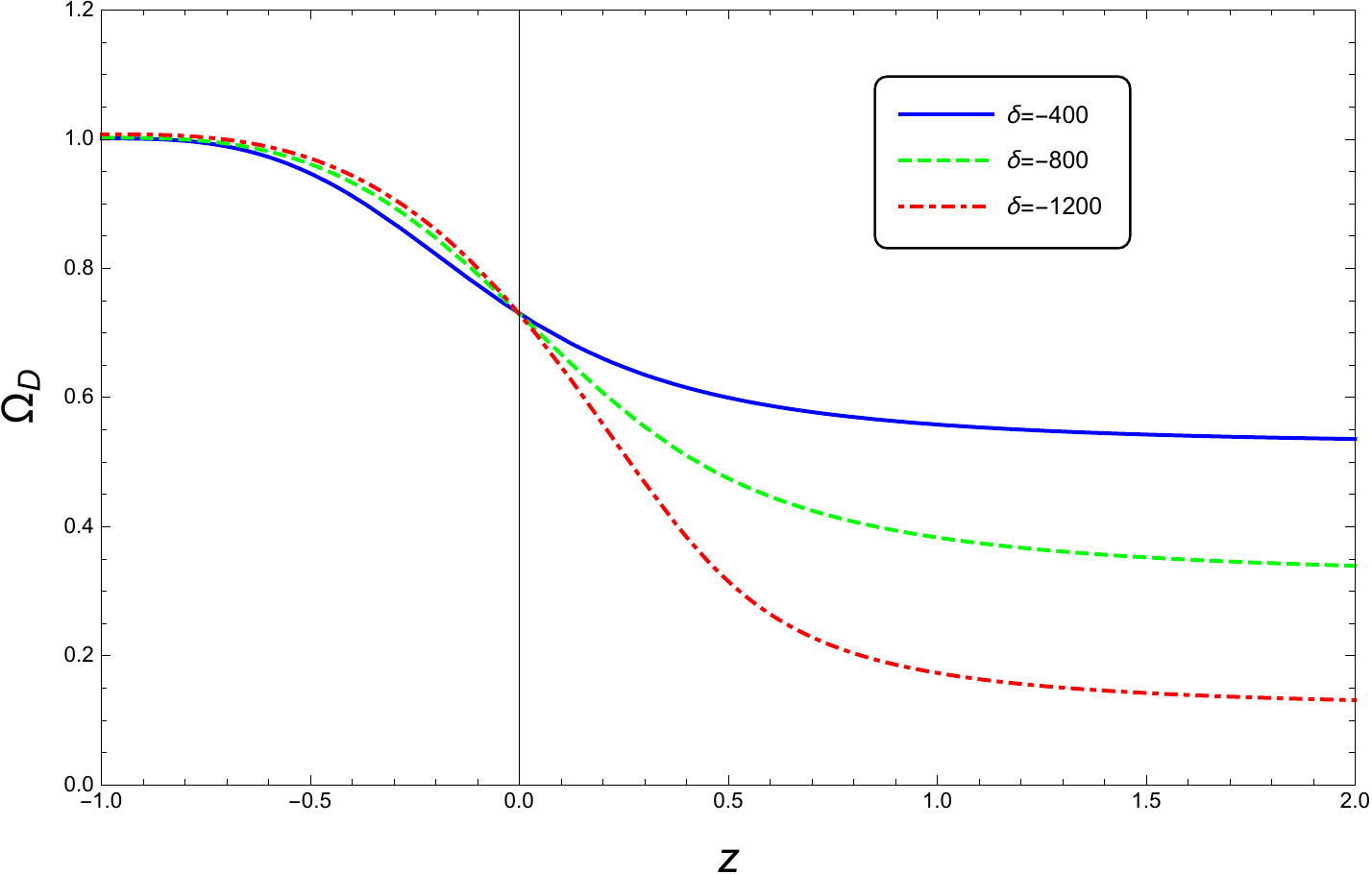}
		\caption{$\Omega_{D}$ vs $z$ for $\Omega_{D0}=0.69$ and different $\delta$ values with interacting fluids.}
		\label{fig:Omd-int}
	\end{subfigure}
	\begin{subfigure}{0.45\textwidth}
		\centering
		\includegraphics[width = 0.8 \linewidth]{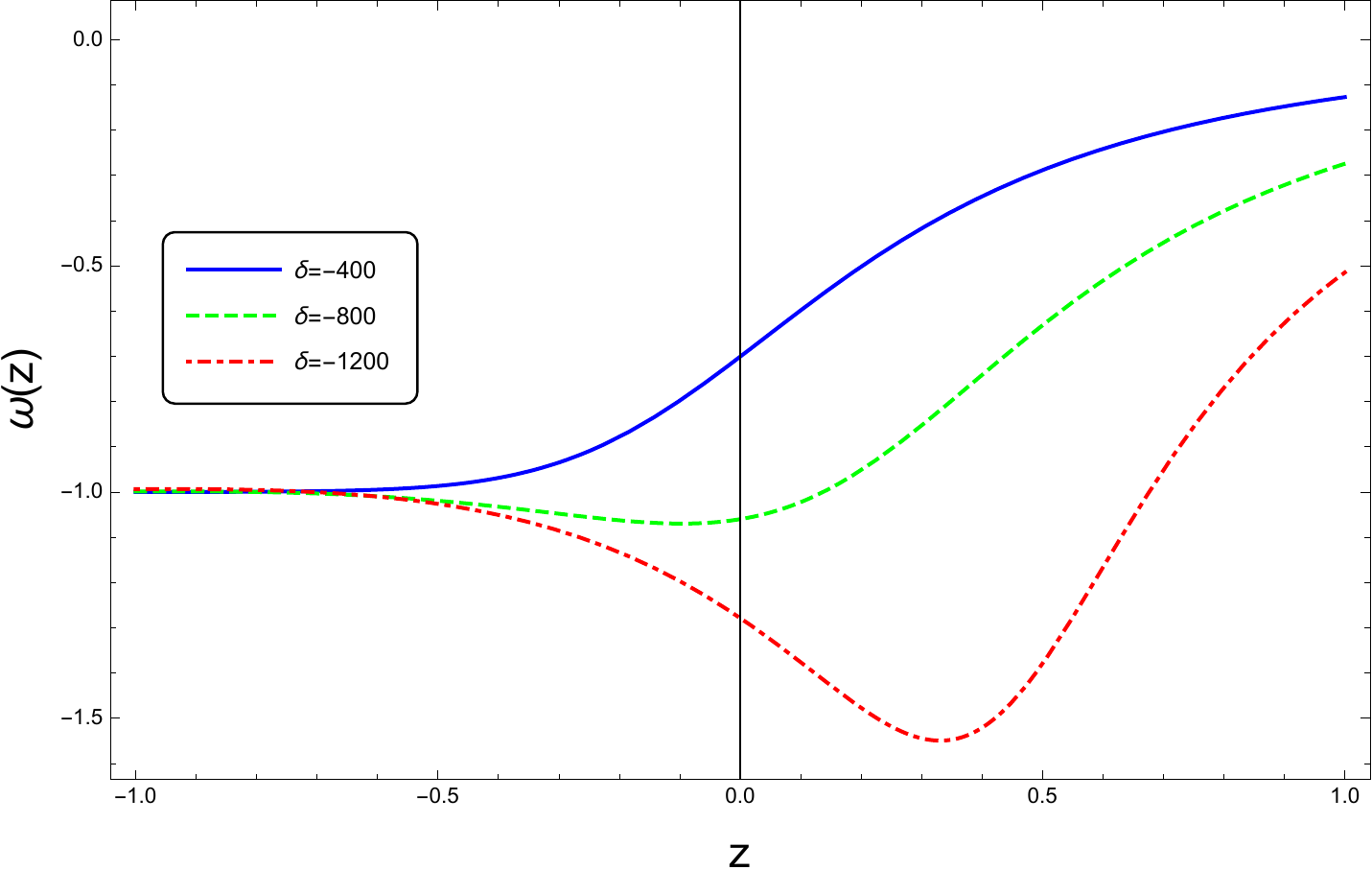}
		\caption{$w(z)$ vs $z$ for $\Omega_{D0}=0.69$ and different $\delta$ values with interacting fluids.}
		\label{fig:eosz-int}
	\end{subfigure}
	\begin{subfigure}{0.45\textwidth}
		\centering
		\includegraphics[width = 0.8 \linewidth]{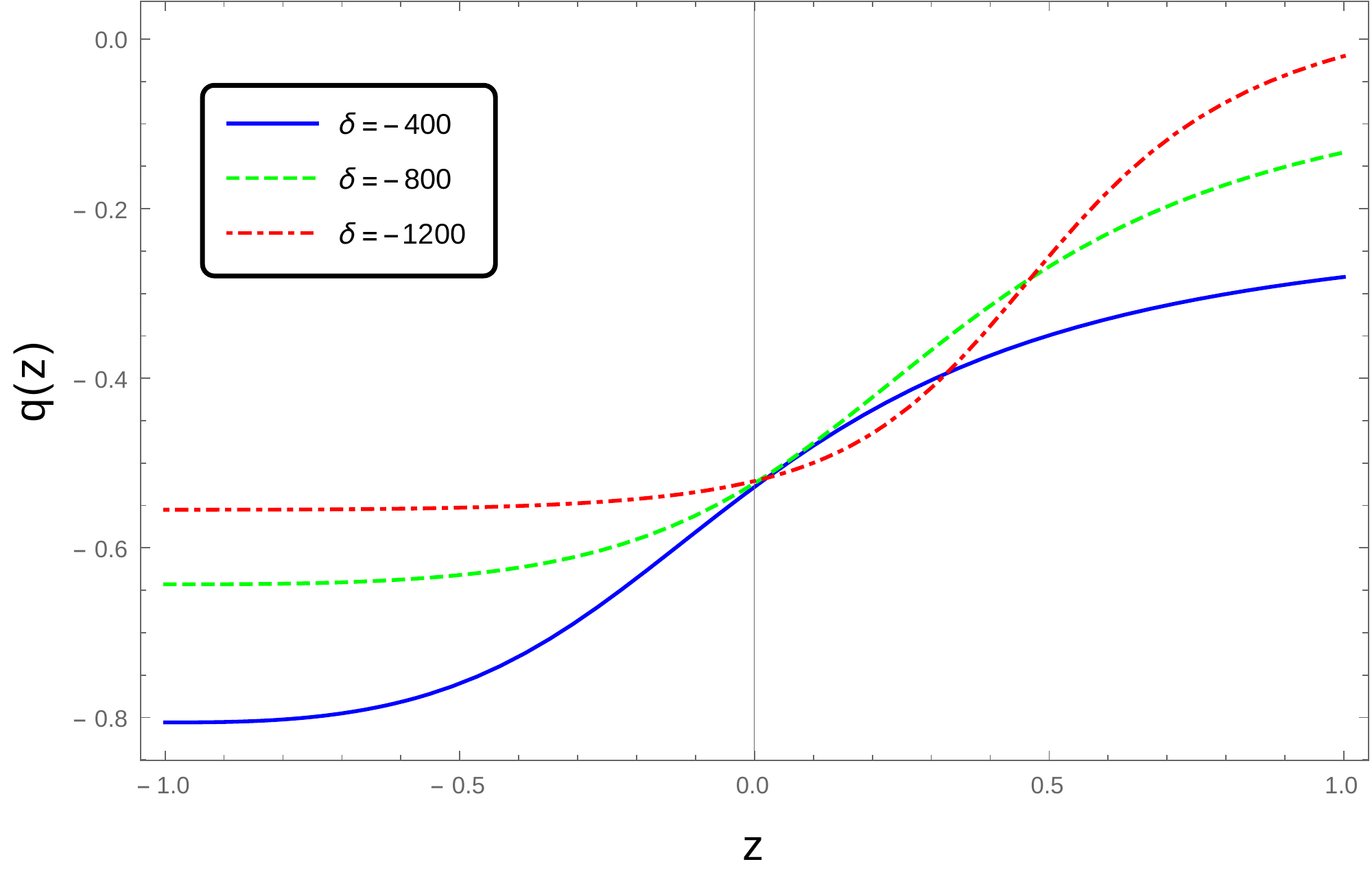}
		\caption{$q(z)$ vs $z$ for $\Omega_{D0}=0.69$ and different $\delta$ values with interacting fluids.}
		\label{fig:decz-int}
	\end{subfigure}	
	\begin{subfigure}{0.45\textwidth}
		\centering
		\includegraphics[width = 0.8 \linewidth]{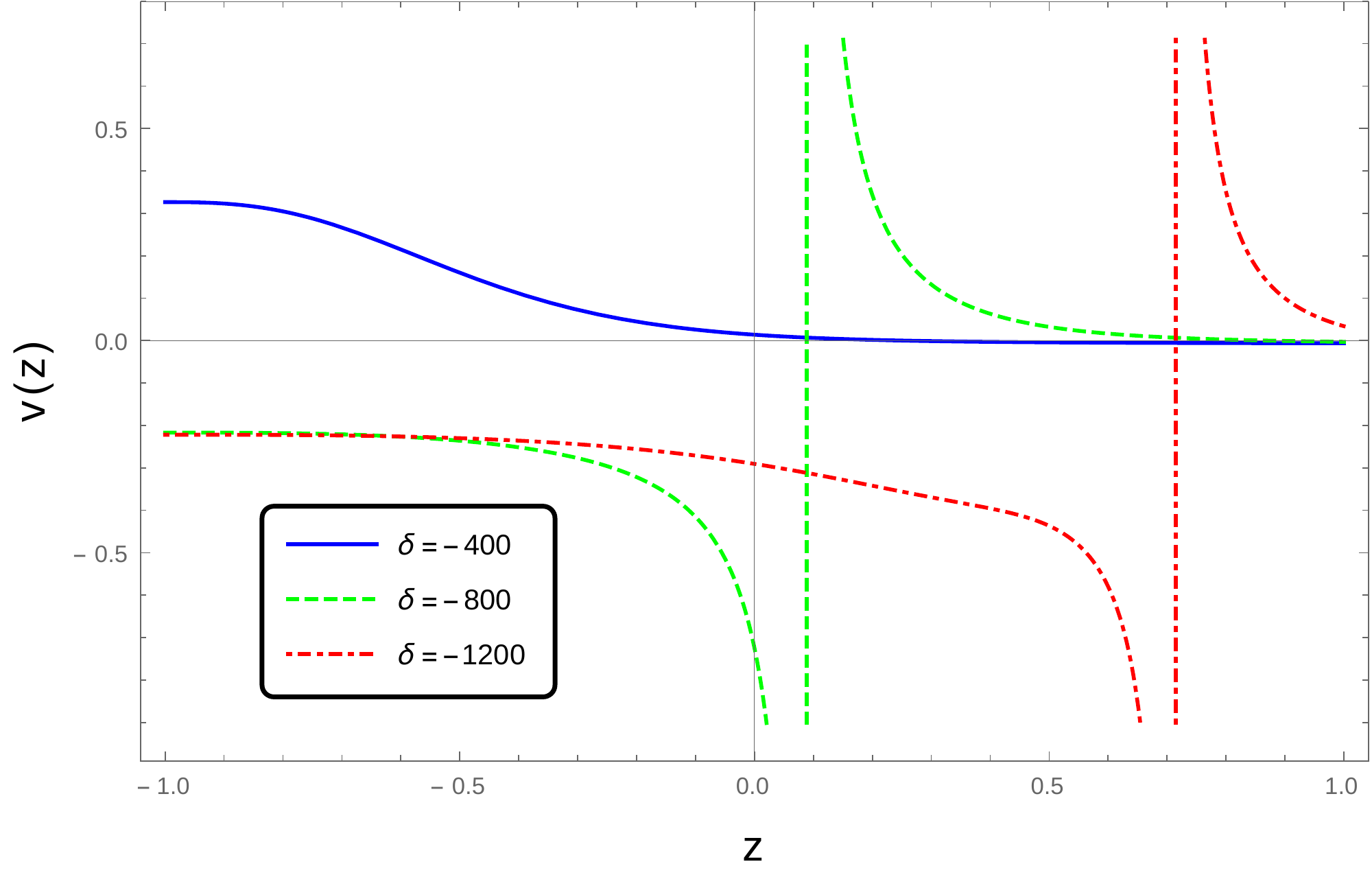}
		\caption{$v^{2}(z)$ vs $z$ for $\Omega_{D0}=0.69$ and different $\delta$ values with interacting fluids.}
		\label{fig:vsq-int}
	\end{subfigure}
	\label{test2}
	\caption{Evolution of cosmological parameters and stability of the interacting RHDE model in DGP brane.}
	
\end{figure}

%*****************************%
\section{RHDE in DGP braneworld with interacting fluids}
\label{rhdeint}
%***************************************%

%***************************************%
In this section, we consider two types of cosmic fluids that are interacting. The evolution of the universe is studied considering interacting DE and matter content present in the universe. The conservation equations, i.e.  eq. (\ref{cons-ni1}) and eq. (\ref{cons-ni2}) are satisfied separately for the non-interacting fluids. But for the interacting model they are related through some interaction:
\begin{equation}
\label{inter1}
    \dot{\rho_{m}}+3H\rho_{m}=Q,
\end{equation}
\begin{equation}
\label{inter2}
    \dot{\rho_{D}}+3H\rho_{D}(1+w_{D})=-Q.
\end{equation}
where $Q$ gives the interaction between the dark energy and dark matter. The total energy density $\rho=\rho_{m}+\rho_{D}$ satisfies the energy conservation equation. We consider the interaction term to be of the form \cite{Sun:2010vz, Wang:2005jx}:
\begin{equation}
\label{interf}
    Q=3b^{2}H(1+r)\rho_{D}=\frac{3b^{2}H(1+2\sqrt{\Omega_{rc}})}{\Omega_{D}}
\end{equation}
where $b$ denotes a coupling constant and $r=\rho_m/\rho_D$. \cite{pavon2005holographic,honarvaryanl} and $r$ is defined as:
\begin{equation}
\label{denratio}
1+r=\frac{1+2\sqrt{\Omega_{rc}}}{\Omega_D}.
\end{equation}
The equation of state parameter, in presence of the interaction, becomes:
\begin{equation}
\label{omd-int} 
w_{D}= \frac{C^{2}(1+3\sqrt{\Omega_{rc}})-\Omega_{D}(1+2\sqrt{\Omega_{rc}})-\frac{b^{2}C^{2}}{\Omega_{D}}(1+2\sqrt{\Omega_{rc}})(1+\sqrt{\Omega_{rc}})}{C^{2}(1+\sqrt{\Omega_{rc}})-(2C^{2}-\Omega_{D})\Omega_{D}}
\end{equation}
Fig.(\ref{fig:Omd-int})-fig.(\ref{fig:decz-int}) illustrate the evolution of different cosmological parameters which remains more or less unchanged from the non-interacting scenario in nature. However, although the nature of evolution is similar in both models the transition redshift etc. differs. A comparison is shown in fig.(\ref{fig:comp}).From eq.(\ref{omprime-ni}) and eq. (\ref{omd-int}) we get:
\begin{equation}
\label{omprime-int}
\begin{aligned}
\Omega'_{D}= & \frac{3(C^{2}-\Omega_{D})\Omega_{D}}{C^{2}(1+\sqrt{\Omega_{rc}})\left[C^{2}(1+\sqrt{\Omega_{rc}})-(2C^{2}-\Omega_{D})\Omega_{D}\right]} \\
& \left[C^{2}(1+3\sqrt{\Omega_{rc}})(1+\Omega_D)+2C^{2}\Omega_{rc}- \right.\\
& \left.2C^2\Omega_D(1+2\sqrt{\Omega_{rc}})-b^{2}C^{2}(1+2\sqrt{\Omega_{rc}})((1+\sqrt{\Omega_{rc}}))\right]. 
\end{aligned}   
\end{equation}
We obtain the expression for deceleration parameter using eq.(\ref{hdot-ni}) and eq. (\ref{omd-int}):
\begin{equation}
\label{decz-int}
\begin{aligned}
q= & \frac{1}{2(1+\sqrt{\Omega_{rc}})[C^{2}(1+\sqrt{\Omega_{rc}})-(2C^{2}-\Omega_{D})\Omega_{D}]}\\
   & \left[(5C^{2}-C^{2}\Omega_{D}-2\Omega_{D}^{2})\sqrt{\Omega_{rc}}+C^{2}(1+4\Omega_{rc})+ \right.\\
	& \Omega_{D}(C^{2}-\Omega_{D})-3\Omega_{D}^{2}-3b^{2}C^{2}(1+2\sqrt{\Omega_{rc}}) \\
	&\left.(1+\sqrt{\Omega_{rc}})\right],
\end{aligned}
\end{equation}
As illustrated in fig.(\ref{fig:decz-int}), in presence of interaction the universe is always accelerating. No phase of deceleration is seen in the past. The present value of the deceleration parameter does depend on the model parameter $\delta$.

\section{Classical Stability of the Models}
\label{clstable}
The square speed of sound is computed to look into the classical stability of the model. For the non interacting case: ($v_{s}^{2}=\frac{dp_{D}}{d\rho_{D}}=\frac{\frac{dp_{D}}{dH}}{\frac{d\rho_{D}}{DH}}$) can be written as:
\begin{equation}
\label{vsq-ni}
\begin{aligned}
v^{2}= & w_{D}+ \frac{C^{2}}{2(2C^{2}-\Omega_{D})\left(C^{2}(1+\sqrt{\Omega_{rc}})-(2C^{2}-\Omega_{D})\Omega_{D}\right)^{2}} \\
& \left[\left((-3C^{2}+2\Omega_{D}) \sqrt{\Omega_{rc}}- \frac{2(C^{2}-\Omega_{D})(1+2\sqrt{\Omega_{rc}})\Omega_{D}}{C^{2}}\right) \right. \\
& \left(C^{2}(1+\sqrt{\Omega_{rc}})-(2C^{2}-\Omega_{D})\Omega_{D}\right)+ \left(C^{2}+3C^{2}\sqrt{\Omega_{rc}} \right.\\
&\left. -\Omega_{D}(1+2\sqrt{\Omega_{rc}})\right)\left. \left((2C^{2}\sqrt{\Omega_{rc}}-\frac{4(C^{2}-\Omega_{D})^{2}\Omega_{D}}{C^{2}})\right)\right].
\end{aligned}
\end{equation}
%****************************%
The evolution of the squared sound speed is shown in fig.(\ref{fig:vsq-ni}). For smaller $\delta$ values the present universe is classically unstable albeit it might have been stable in some past epoch. Larger $\delta$ values can give a stable universe in the present epoch but as we have seen that costs us an earlier phase of deceleration. For smaller values such as $\delta = -800$ there are regions where the analysis fails as the sound speed diverges. The adiabatic sound speed for the interacting scenario is given by:
\begin{equation}
\label{vsq-in}
\begin{aligned}
v^{2}= & w_{D}+ \frac{c^{2}}{2(2C^{2}-\Omega_{D})\left(C^{2}(1+\sqrt{\Omega_{rc}})-(2C^{2}-\Omega_{D})\Omega_{D}\right)^{2}} \\
       & \left[\left((-3C^{2}+2\Omega_{D})+\frac{b^{2}C^{2}}{\Omega_D}(3+4\sqrt{\Omega_{rc}})\sqrt{\Omega_{rc}}\right)- \right.\\
			&\frac{2(C^{2}-\Omega_{D})(1+2\sqrt{\Omega_{rc}})\Omega_{D}}{C^{2}}\left(1- \frac{b^{2}C^{2}}{\Omega_D}(1+\sqrt{\Omega_{rc}})\right)  \\
			& \left(C^{2}(1+\sqrt{\Omega_{rc}})-(2C^{2}-\Omega_{D}\Omega_{D})\right)+ \left(C^{2}+3C^{2}\sqrt{\Omega_{rc}}-\Omega_{D}(1+2\sqrt{\Omega_{rc}}) \right. \\
& \left.\left. -\frac{b^{2}C^{2}}{\Omega_D}(1+2\sqrt{\Omega_{rc}})(1+\sqrt{\Omega_{rc}})\right) \left((2C^{2}\sqrt{\Omega_{rc}}-\frac{4(C^{2}-\Omega_{D})^{2}\Omega_{D}}{C^{2}})\right)\right].
\end{aligned}    
\end{equation}
Evolution of $v^2$ for the interacting case is shown in fig.(\ref{fig:vsq-int}). Similar to the non-interacting case, we find that the model is classically unstable in the present epoch but was stable in the past. There are regions where the sound speed diverges and the analysis fails. Note that for a larger value $\delta$ one may obtain a stable present phase in the interacting model also but then again we no longer obtain a phase of deceleration.

\section{Diagnostic Tests}
\label{sec:diagn}
\begin{figure}[ht!]
	\centering
	\includegraphics[width = 0.7 \linewidth]{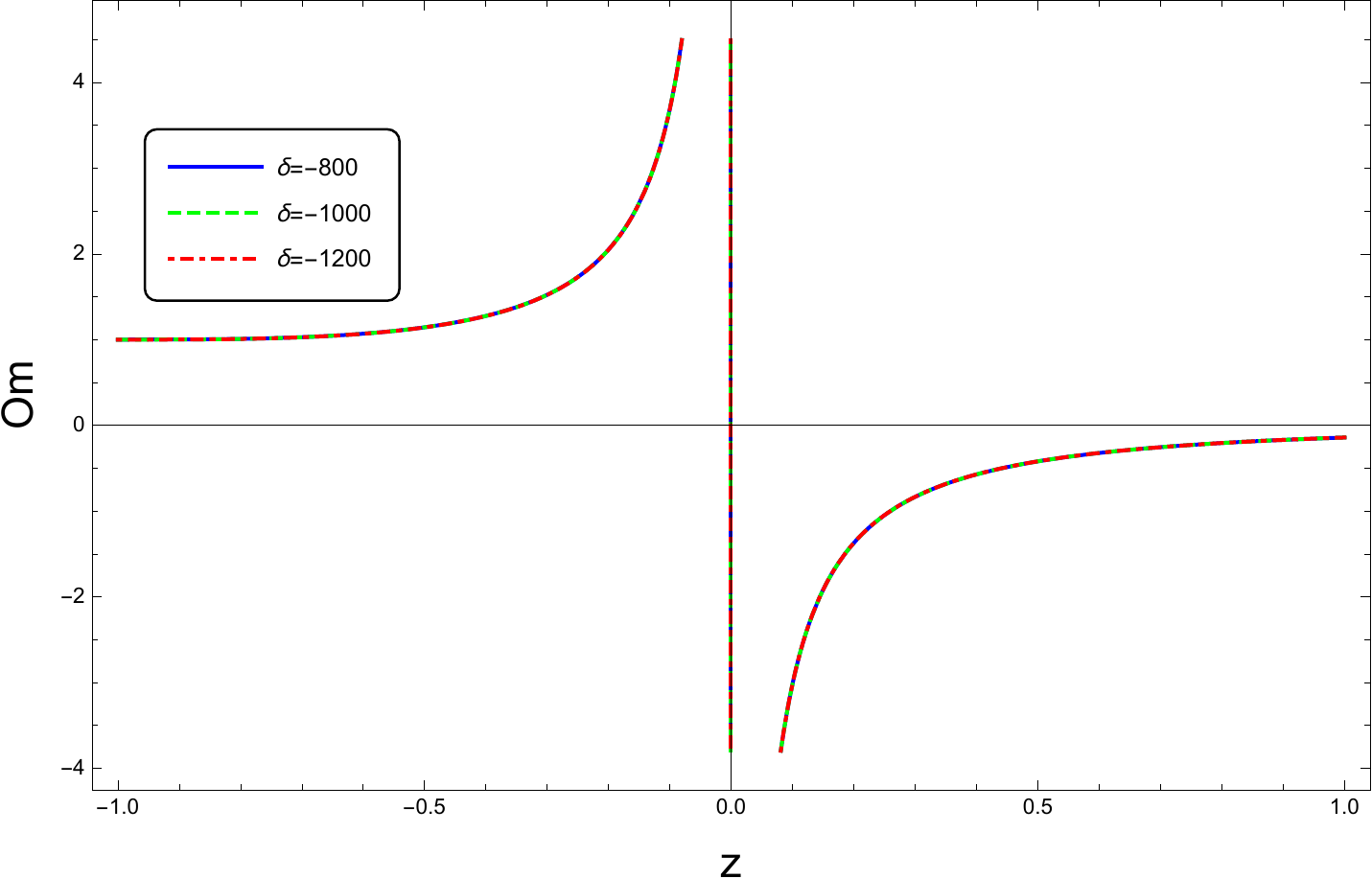}
	\caption{Om-diagnostic of the models($Om(z)$ vs $z$ ) for different $\delta$ values.}
	\label{fig:omdz}
\end{figure}

\subsection{\emph{Om} Diagnostic}
\label{sec:omd}
We study Om diagnostic of the models to explore the nature of dark energy in different epochs of evolution. Om diagnostic has an additional advantage of differentiating dynamical dark energy models from the $\Lambda$CDM model \cite{shahalam2015om}. For a spatially flat universe, $Om(z)$ is defined as:
\begin{equation}
\label{omd}
Om(z) = \frac{H^2/H^2_{0}-1}{(1+z)^3-1}.
\end{equation}
Om being geometric in nature presents three distinct scenario: (i) a positive slope of $Om(z)$ indicates a phantom nature of the dark energy (ii) a negative slope indicates quintessence, and (iii) a constant $Om(z)$ signifies the $\Lambda$CDM nature. Fig. (\ref{fig:omdz}) shows our findings. All the above phases appear in the course of evolution. The diagnostics of both models overlap for different values of $\delta$.
\subsection{Statefinder Diagnostic} 
\label{sec:sf}  
\begin{figure}[ht!]
	\centering
	\begin{subfigure}{0.45\textwidth}
		\centering
		\includegraphics[width = 0.8 \linewidth]{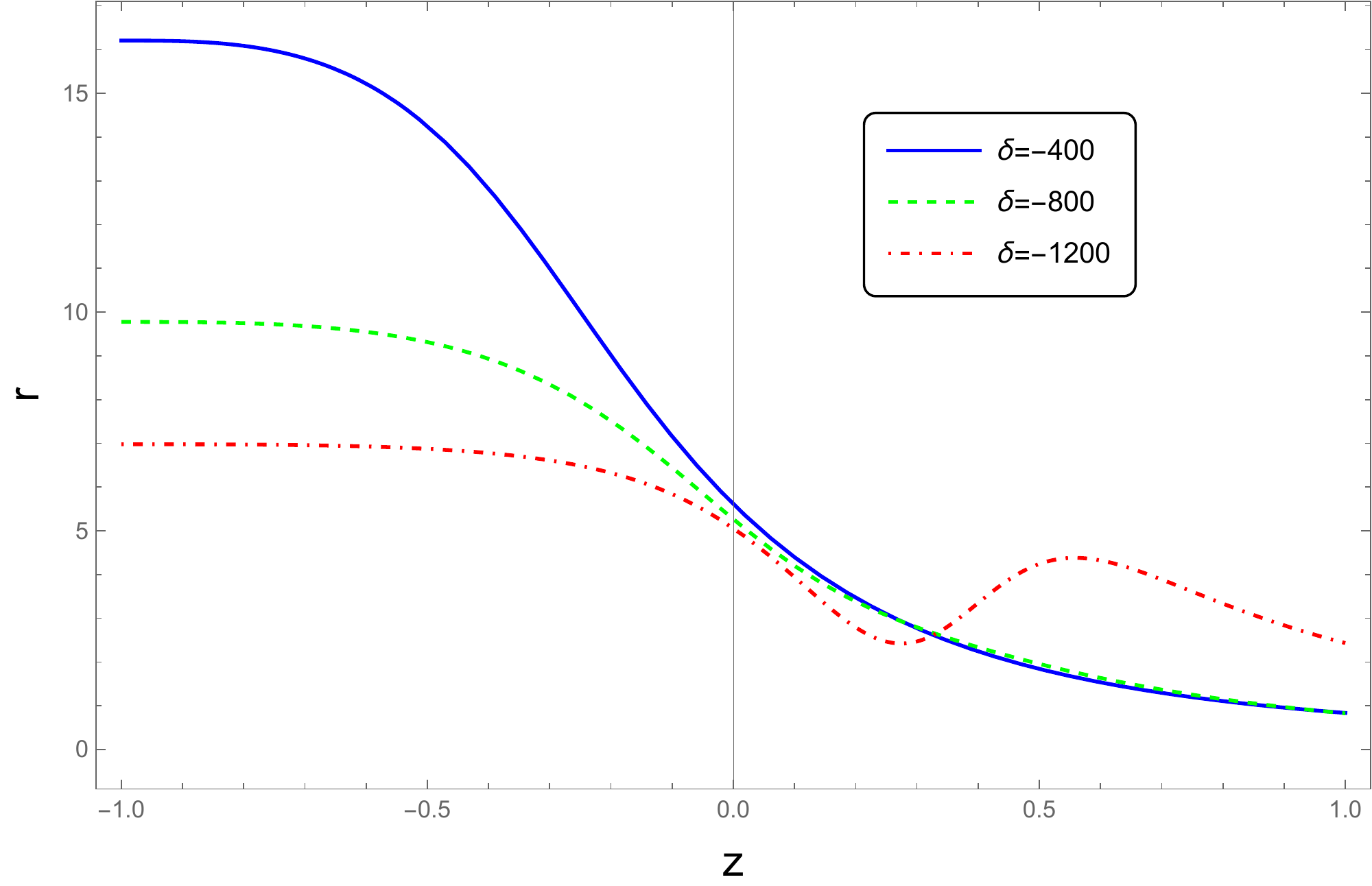}
		\caption{$r(z)$ vs $z$ in non-interacting scenario.}
		\label{rnonint}
	\end{subfigure}
	\begin{subfigure}{0.45\textwidth}
		\centering
		\includegraphics[width = 0.8 \linewidth]{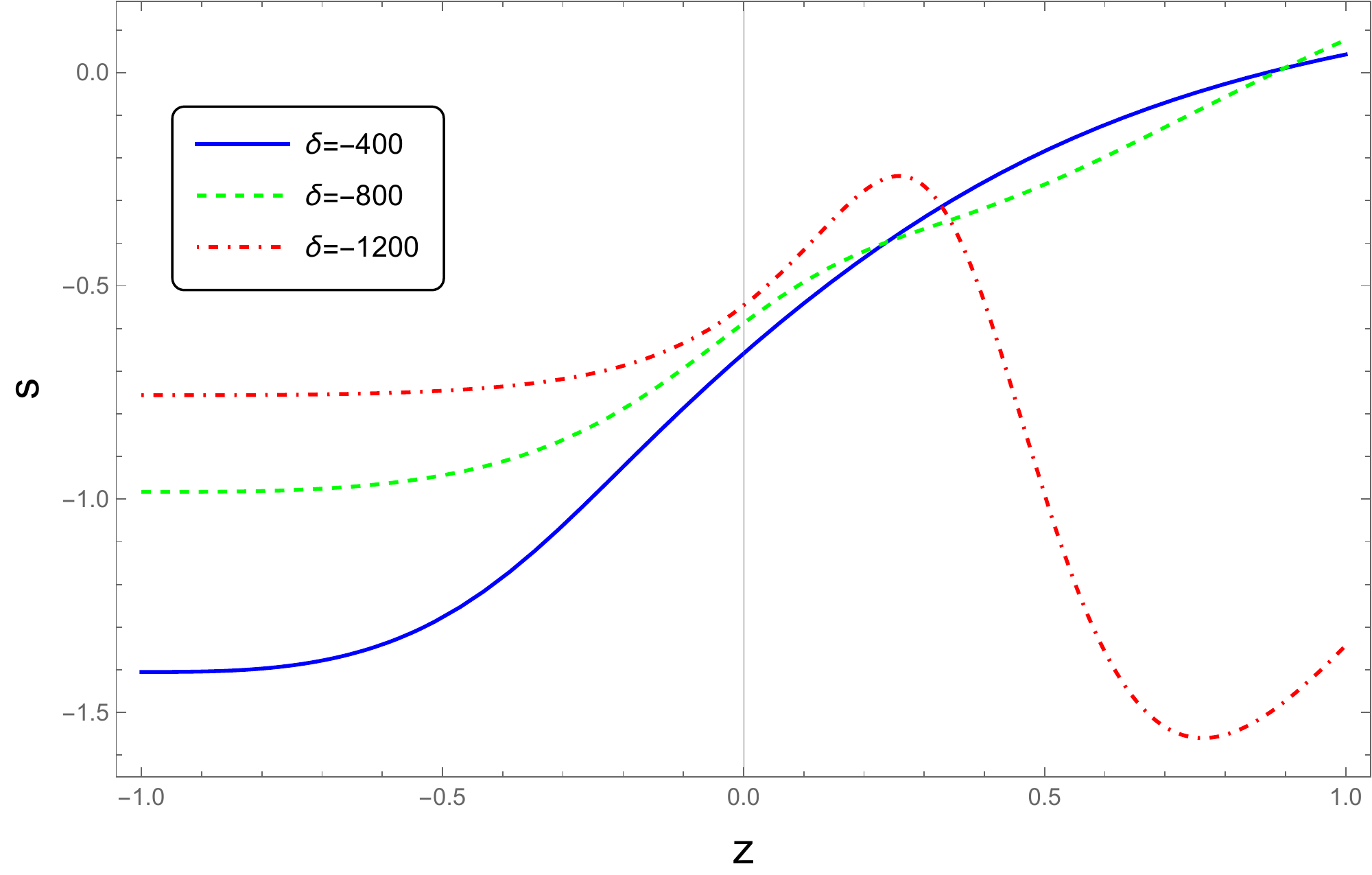}
		\caption{$s(z)$ vs $z$ in non-interacting scenario.}
		\label{snonint}
	\end{subfigure}
	\begin{subfigure}{0.45\textwidth}
		\centering
		\includegraphics[width = 0.8 \linewidth]{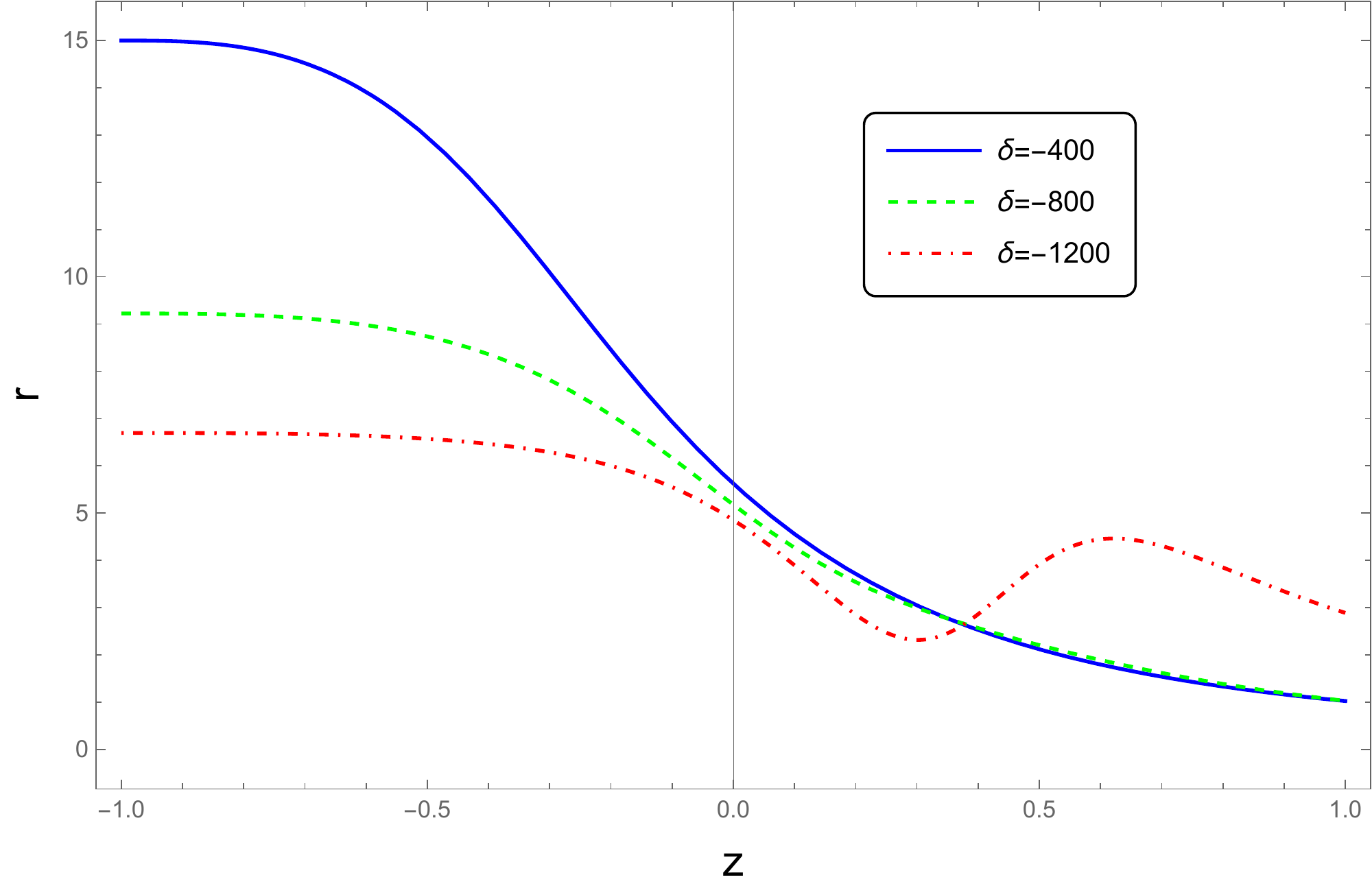}
		\caption{$r(z)$ vs $z$ in interacting fluid scenario.}
		\label{rint}
	\end{subfigure}	
	\begin{subfigure}{0.45\textwidth}
		\centering
		\includegraphics[width = 0.8 \linewidth]{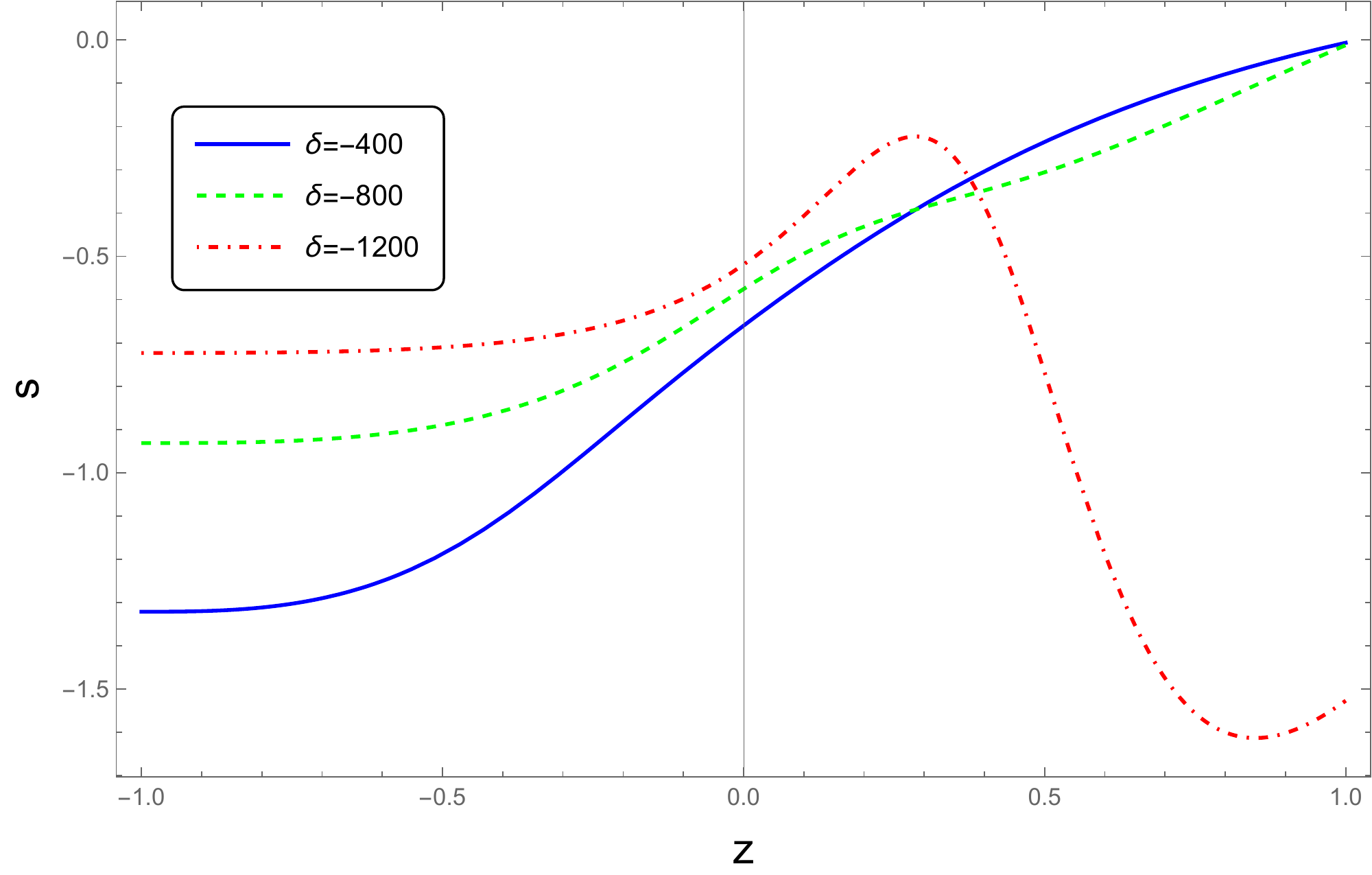}
		\caption{$s(z)$ vs $z$ in interacting fluid scenario.}
		\label{sint}
	\end{subfigure}
	\label{rs}
	\caption{Evolution of statefinder parameters in case of non-interacting and interacting RHDE model in DGP brane.}
\end{figure}
To distinguish between different cosmological models Sahni ${\it et. al.}$ \cite{Sahni:2002fz} introduced a geometric diagnostic tool known as the statefinder diagnostics. The statefinder pair $\{r,s\}$ is derived from the scale factor $a$ and is given by,
\begin{equation}
\label{rs}
r=2q^{2}+q-\frac{\dot{q}}{H},\;\;\;\;\;\;\;s=\frac{r-1}{3(1-\frac{1}{2})},
\end{equation}
with $q=-\frac{a\ddot{a}}{\dot{a}^2}$ being the deceleration parameter. 
Statefinder diagnostic is a geometrical diagnostic tool which depends on the scale factor and thus on the underlying metric structure. Based on the trajectories of the statefinder pair one can distinguish between the $\Lambda$CDM cosmology and other DE scenarios like quintessence, phantom etc. For $\Lambda$CDM the pair corresponds to $\{r,s\}=\{1,0\}$. In case of interacting RHDE model the deceleration parameter can be recasted as,
\begin{equation}
\label{qnew}
q=\frac{A-B}{2(1+\sqrt{\Omega_{rc}})D},
\end{equation}
where, $A=(5C^{2}-C^{2}\Omega_{D}-2\Omega_{D}^{2})\sqrt{\Omega_{rc}}+C^{2}(1+4\sqrt{\Omega_{rc}})+\Omega_{D}(C^{2}-\Omega_{D})-3\Omega_{D}^{2}$, $B=-3b^{2}C^{2}(1+2\sqrt{\Omega_{rc}})(1+\sqrt{\Omega_{rc}})$ and $D=C^{2}(1+\sqrt{\Omega_{rc}})-(2C^{2}-\Omega_{D})\Omega_{D}$. Also,
\[\frac{\dot{q}}{H}=\frac{\sqrt{\Omega_{rc}}}{2(1+\sqrt{\Omega_{rc}})^{2}}\Big[1+4\sqrt{\Omega_{rc}}+\]
\begin{equation}
\label{qdot}
3\omega_{D}\Omega_{D}\Big]\frac{\dot{H}}{H^{2}}+\frac{1}{2(1+\sqrt{\Omega_{rc}})}\Big[-4\sqrt{\Omega_{rc}}\frac{\dot{H}}{H^{2}}+3\omega_{D}'\Omega_{D}+3\omega_{D}\Omega_{D}'\Big],
\end{equation}
with,
\begin{equation}
\label{wdp}
\omega_{D}'=\frac{1}{D^{2}}\Big(Da_{1}+a_{2}\Big)\frac{\dot{H}}{H^{2}},
\end{equation}
where, $a_{1}=\Big[(-3C^{2}+2\Omega_{D})+\frac{b^{2}C^{2}}{\Omega_{D}}(3+4\sqrt{\Omega_{rc}})\Big]\sqrt{\Omega_{rc}}-\frac{2(C^{2}-\Omega_{D})(1+2\sqrt{\Omega_{rc}})\Omega_{D}}{C^{2}}\Big[1-\frac{b^{2}C^{2}}{\Omega_{D}^{2}}(1+\sqrt{\Omega_{rc}})\Big]$ and $a_{2}=\Big[C^{2}(1+3\sqrt{\Omega_{rc}})-\Omega_{D}(1+2\Omega_{rc})-\frac{b^{2}C^{2}}{\Omega_{D}}(1+2\sqrt{\Omega_{rc}})(1-\sqrt{\Omega_{rc}})\Big]\Big[2C^{2}\sqrt{\Omega_{rc}}-\frac{4(C^{2}-\Omega_{D})}{C^{2}}\Omega_{D}\Big]$.\\
The evolution of the statefinder pair $r(z)$ and $s(z)$ are shown in Figs. (\ref{rnonint}), (\ref{snonint}), (\ref{rint}) and (\ref{sint}) for both non-interacting and interacting scenarios. From the figures we note that for both the interacting as well as non-interacting scenarios the model differs from the standard $\Lambda$CDM cosmology at the present epoch. This behaviour is independent of the choice of $\delta$. The evolutionary curves lie in the $r>1$, $s<0$ domain and there is a significant deviation from the $\Lambda$CDM region at the present epoch and near future for the chosen $\delta$ values.

\section{Discussion}
\label{disc}
%******************************************************

\begin{figure}[ht!]
	\centering
	\begin{subfigure}{0.45\textwidth}
		\centering
		\includegraphics[width = 0.8 \linewidth]{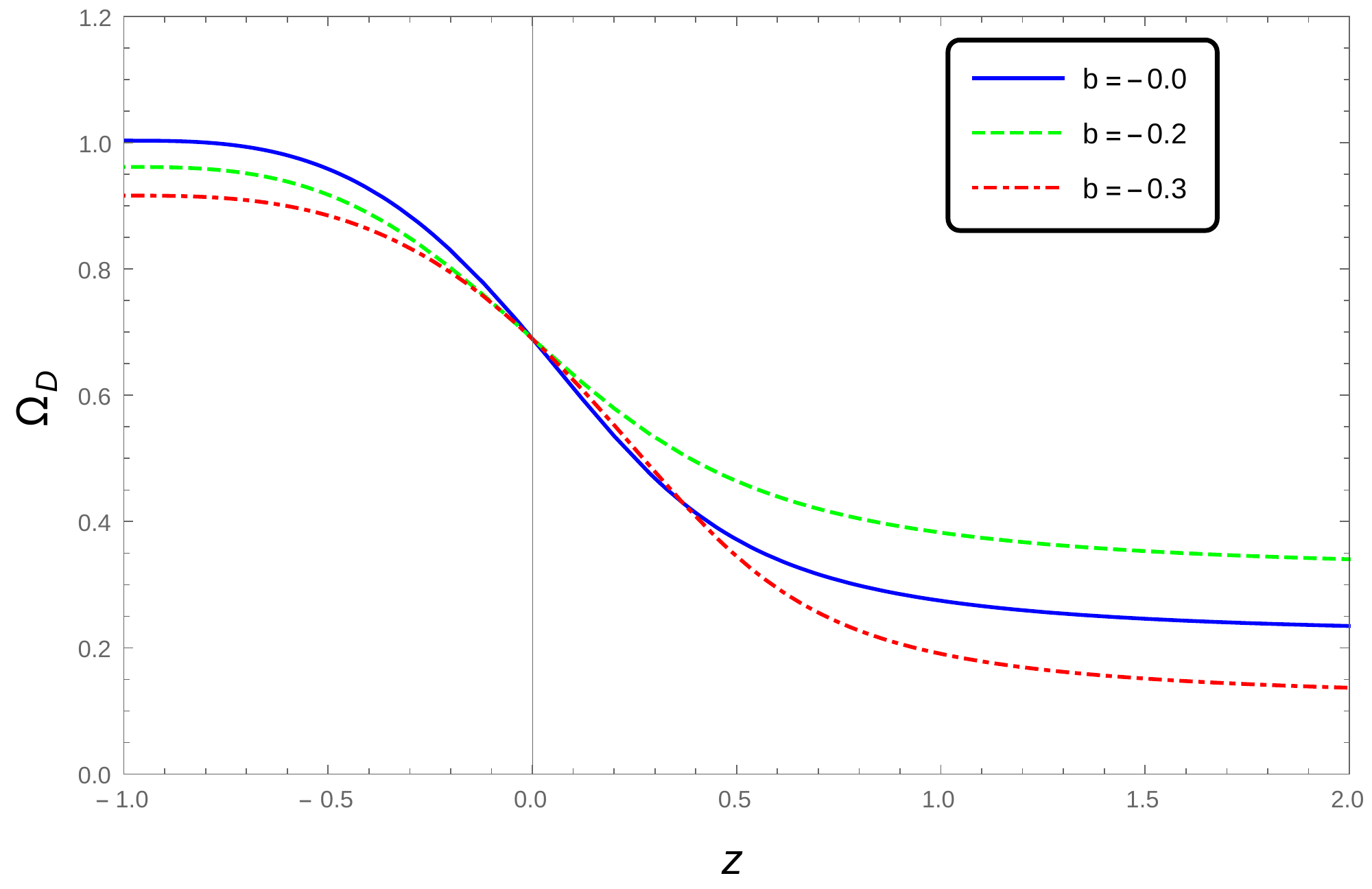}
		\caption{$\Omega_D$ vs $z$}
		\label{fig:Omd-com}
	\end{subfigure}
	\begin{subfigure}{0.45\textwidth}
		\centering
		\includegraphics[width = 0.8 \linewidth]{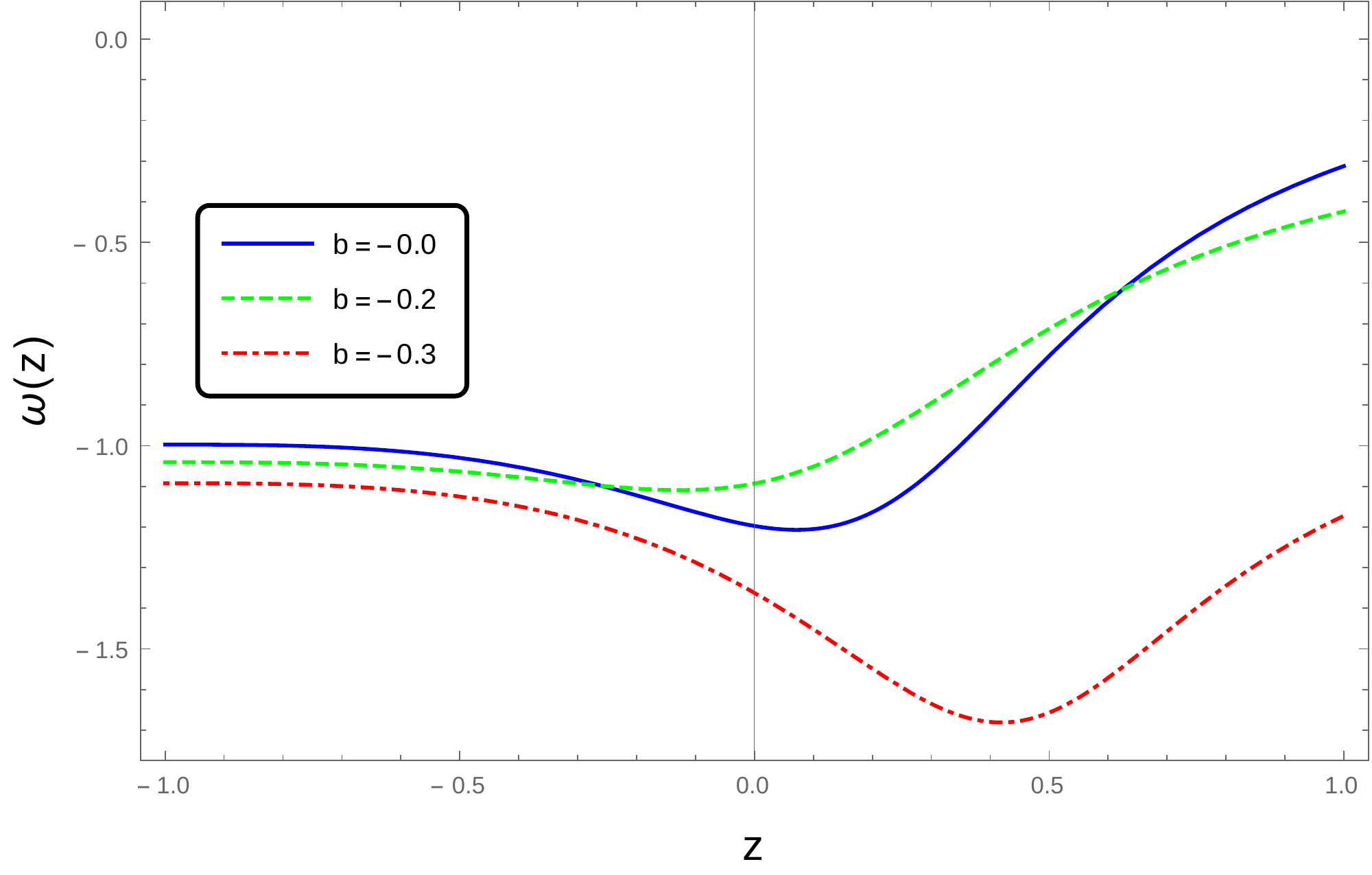}
		\caption{$\omega_D$ vs $z$}
		\label{fig:eosz-com}
	\end{subfigure}
	\begin{subfigure}{0.45\textwidth}
		\centering
		\includegraphics[width = 0.8 \linewidth]{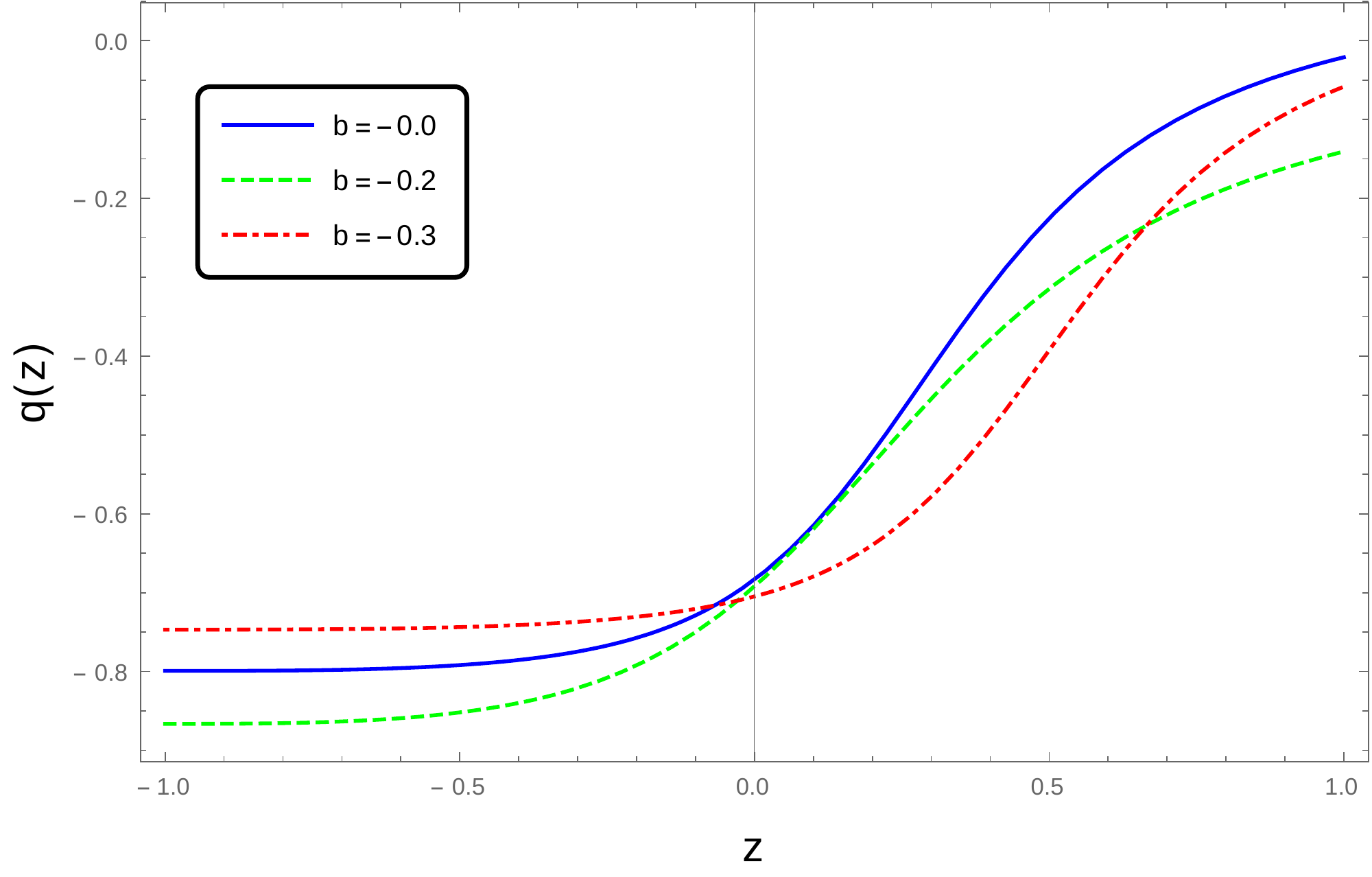}
		\caption{$q(z)$ vs $z$}
		\label{fig:decz-com}
	\end{subfigure}	
	\caption{comparison of Interacting and Non-interacting Models for $\delta = -1000$}%
	\label{fig:comp}
	
\end{figure}
%*****************************************************
We present cosmological models with dark energy described by the RHDE  in the DGP brane-world cosmology with and without interaction among the cosmic fluids in the universe. Both models accommodate the present accelerating phase and share a lot of common features. It is found that both the cosmologies accommodate the present phase of dark energy that is effectively phantom fluid in nature. For large values of $\delta$ ($\delta > -400$), the dark energy, however, does not enter into a phantom phase of expansion. It is found that the dark energy density parameter ($\Omega_D$) increases with time in both cases but converges in the future epoch. Both the cosmologies are found to satisfy the observed value $\Omega_D$ ($\Omega_D\approx 0.69$) which are consistent with the recent observations such as the Planck \cite{aghanim2020planck}.  For different $\delta$ values the EOS parameters ($\omega_D$) are found to converge in future, where $\omega_D\approx-1$), although they differ substantially in the past. We do not find any cosmological model that accommodates a past phase of deceleration for higher $\delta$ values ($\delta \geq -800$). Both models suffer from the same problem. It is difficult to envisage an eternally accelerating universe since an early decelerating one is more suited to structure formation. Similarly, positive $\delta$ values correspond to the nonphysical evolution of cosmological parameters. Although the present value of the deceleration parameter is almost the same in the presence or absence of interaction having a wide range of $\delta$ values, it is also found that $q$ significantly differs in the past and future evolution.  Although the nature of evolution is found to be similar in both models important phases of evolution such as transition, and redshift are found different in interacting as well as in non-interacting fluids even for the same $\delta$ value (see fig.(\ref{fig:comp})).
\begin{figure}[ht!]
	\centering
	\includegraphics[width= 0.7\textwidth]{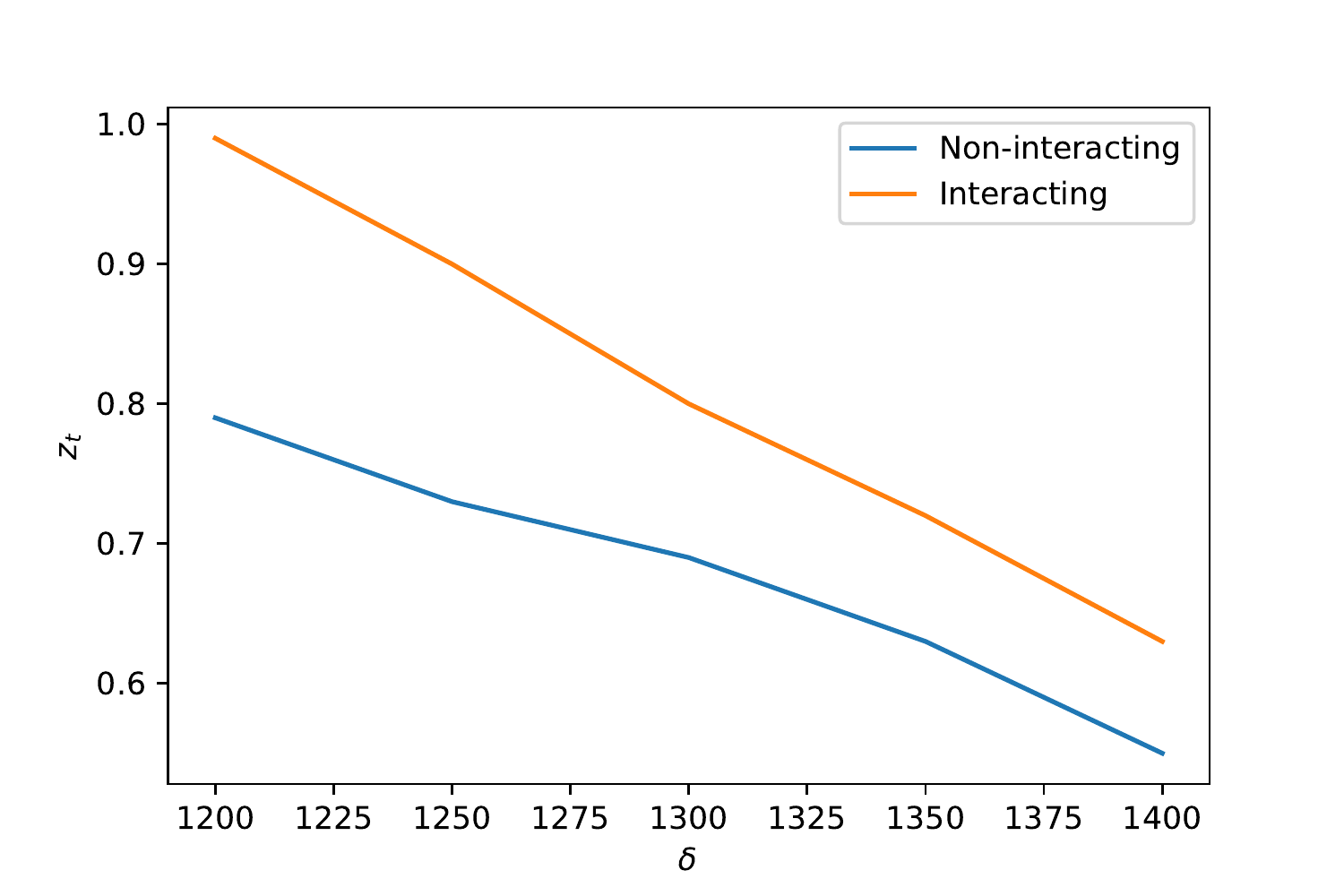}
	\caption{Variation of transition redshift ($z_t$) with $\delta$.}
	\label{fig:transit}
\end{figure}
Fig. (\ref{fig:transit}) presents the variation of transition redshift with the model parameter $\delta$ for both interacting and non-interacting case. Both scenarios can accept the presently accepted value for transition redshift ($z_t < 1$) \cite{Kumar:2022mtx}. For example values like $z_t = 0.69^{+0.25}_{-0.25}$ as suggested in \cite{Velasquez} can be incorporated. Note that, for the same value of $\delta$ the transition occurs at an earlier epoch in the interacting model. The cosmological models are found to transit to $\Lambda$CDM ($\omega_D = -1$) scenario in future. This is also evident from the $Om$-diagnostic analysis of the models presented in \ref{sec:omd}. The $Om$ diagnostic also points out that the dark energy evolves through both the phantom and quintessence phases. We found that the $Om$ diagnostic is insensitive to the model parameter $\delta$ and consequently performed the statefinder diagnostic in sec. (\ref{sec:sf}).  None of the models, though, is classically stable against small perturbation at present for smaller $\delta$ (eg. $\delta = -1000$). For relatively larger values of $\delta$ (eg. $\delta = -400$) both models are classically stable throughout. The standard way of implementing RHDE has been criticized in a very recent article \cite{Manoharan:2022qll} where a proposal for R\'{e}nyie modified Friedmann equations has been put forward. Although the \cite{Manoharan:2022qll} is focused on standard GR, their suggestion might have a general applicability.  It would be thus interesting to explore what constraint recent observations put on the model parameter $\delta$, thus further clarifying the viability of these models. Recently, many holographic dark energy models were tasted using Supernovae data, Baryon Acoustic Oscillations and cosmic chronometers, Cosmic Microwave Background Radiation, Gamma Ray Burst etc.  \cite{Hernandez, Anagnostopoulos:2020ctz, Sadri:2019qxt}. Full scale data analysis is, however, beyond the scope of the present work and would be taken up elsewhere.
\section*{Acknowledgement}
Authors would like to thank the anonymous referees whose remarks helped immensely in improving the manuscript. SG would like to thank Prof. Hooman Moradpour, Research Institute for Astronomy and Astrophysics of Maragha (RIAAM), Iran for valuable discussion.
\section*{Declaration}

The authors have no relevant financial or non-financial interests to disclose.

\section*{Data availability statement}
There is no data associated with this work.

%\bibliographystyle{ws-mpla}
%\bibliography{dgpref}
\end{document}